\documentclass[prx,twocolumn,superscriptaddress,floatfix,nopacs,nofootinbib]{revtex4-2}
\usepackage{graphicx,amsfonts,amssymb,amsmath,hyperref,enumerate}
\usepackage{algorithm}
\usepackage{algpseudocode}
\usepackage{pgfplots}
\usepackage{pgf}
\usepackage{lmodern}
\usepackage{import}
\usepackage{xr}
\usepackage{enumitem}
\usepackage{soul}
\usepackage{caption}
\usepackage{subcaption}

\newif\ifhyper
\hypertrue
\ifhyper
\hypersetup{
   citecolor = {green},
   colorlinks = {true}, % false
   urlcolor = {blue} % magenta
}
\fi

\newcommand{\beq}{\begin{equation}}
\newcommand{\eeq}{\end{equation}}
\newcommand{\beqa}{\begin{eqnarray}}
\newcommand{\eeqa}{\end{eqnarray}}

\newcommand{\comment}[1]{}

\def\Longarrow{\protect\@lra}
\def\@lra{\relbar\joinrel\relbar\joinrel\relbar\joinrel%
          \relbar\joinrel\rightarrow}

\begin{document} 

\title{Financial Index Tracking via Quantum Computing with Cardinality Constraints}

\author{Samuel Palmer}
\affiliation{Multiverse Computing, Centre for Social Innovation, 192 Spadina Avenue Suite 509, Toronto, ON M5T 2C2, Canada}

\author{Konstantinos Karagiannis}
\affiliation{Quantum Computing Services, Protiviti, 2884 Sand Hill Rd \# 200, Menlo Park, CA 94025, USA}

\author{Adam Florence}
\affiliation{Advanced Analytics, Ally Financial, Ally Charlotte Center, 601 S Tryon St, Charlotte, NC 28202, USA}

\author{Asier Rodriguez}
\affiliation{Multiverse Computing, Parque Cientifico y Tecnol\'{o}gico de Gipuzkua, Paseo de Miram\'{o}n, 170 3$^{\,\circ}$ Planta, E-20014 San Sebasti\'{a}n, Spain}

\author{Rom\'{a}n Or\'{u}s}
\affiliation{Multiverse Computing, Parque Cientifico y Tecnol\'{o}gico de Gipuzkua, Paseo de Miram\'{o}n, 170 3$^{\,\circ}$ Planta, E-20014 San Sebasti\'{a}n, Spain}
\affiliation{Donostia International Physics Center, Paseo Manuel de Lardizabal 4, E-20018 San Sebasti\'{a}n, Spain}
\affiliation{Ikerbasque Foundation for Science, Maria Diaz de Haro 3, E-48013 Bilbao, Spain}

\author{Harish Naik}
\affiliation{Advanced Analytics, Ally Financial, Ally Charlotte Center, 601 S Tryon St, Charlotte, NC 28202, USA}

\author{Samuel Mugel}
\affiliation{Multiverse Computing, Centre for Social Innovation, 192 Spadina Avenue Suite 509, Toronto, ON M5T 2C2, Canada}

\begin{abstract}
In this work, we demonstrate how to apply non-linear cardinality constraints, important for real-world asset management, to quantum portfolio optimization. This enables us to tackle non-convex portfolio optimization problems using quantum annealing that would otherwise be challenging for classical algorithms. Being able to use cardinality constraints for portfolio optimization opens the doors to new applications for creating innovative portfolios and exchange-traded-funds (ETFs). We apply the methodology to the practical problem of enhanced index tracking and are able to construct smaller portfolios that significantly outperform the risk profile of the target index whilst retaining high degrees of tracking.
\end{abstract}

\maketitle

\section{Introduction}

Financial index tracking is an essential application of portfolio optimization. This is  used by financial firms for asset management strategies, or for creating and managing new financial products such as exchange-traded funds (ETFs). Financial indexes often consist of hundreds or thousands of assets. An important task to make it practical to manage such portfolios is replicating the financial index using a limited subset of assets, known as cardinality constraints. Such a constrained portfolio optimization is extremely difficult to solve, and standard algorithms tend to have a very hard time in finding appropriate solutions.   

In this work we propose a way around this issue by using quantum computing. In particular, we use quantum annealing to construct exact cardinality constrained portfolios to allow for practical, real-world management of financial index tracking portfolios. Our solutions are built using quantum computers that are commercially available today.

This paper is organized as follows: in Sec.\ref{port} we refresh how quantum computing can be used in the context of the portfolio optimization problem. In Sec.\ref{card} we explain the cardinality constraint, and how this can be implemented in the quantum computer. Sec.\ref{ind} deals with index tracking, where we discuss also the data setup as well as the obtained results. In Sec.\ref{enh} we explain how to devise a strategy for enhanced index tracking together with results. 

\section{Quantum Portfolio Optimization}
\label{port}

The Markowitz portfolio optimization problem was first introduced for quantum computing by Rosenberg \emph{et al} \cite{Rosenberg2016}. In that paper, the authors noted that they could only solve small problems, and that future hardware would allow for more complex problems and better success rates. We found that this expected progression has arrived with the availability of D-Wave's modern Hybrid Solver annealer, but to get full performance benefits we are required to look at more practically relevant and complex problems. This puts the focus on the problem formulation which is the central point of this paper.

In the Rosenberg paper the problem is formulated as the discrete portfolio optimization. This discrete portfolio optimization problem is known to be non-convex and NP-Hard and can be used as an approximation to the continuous case to be solved on a quantum device. Previous work on quantum portfolio optimization has optimized portfolios of 40, 60, 100 and 500 assets \cite{Cohen2020, Cohen2020a, Palmer2021, Certo2022}, as well as introduced additional constraints such as minimum holding period, investment band, group holdings and volatility constraints \cite{Mugel2021, Palmer2021,Certo2022}.

Quantum portfolio optimization relies on transforming the problem into a quadratic unconstrained binary optimization (QUBO) that can be embedded as interactions between a set of quantum bits (qubits) on a quantum computer. The key ingredient is encoding the asset weights as a set of binary variables. If we have a discrete number of units of investments, $K$, then using binary encoding the weight of an asset $\omega_i$ can be given by

\begin{align}
    \omega_i &= \frac{1}{K}\sum_d^D2^dx_{i,d} \text{ ; } ~~ x_{i,d} \in \mathbb{Z}_2
\end{align}

where $x_{i,d}$ are the binary asset holding variables, and $D$ is the total number of variables per asset, where $\sum_d^D2^dx_{i,d} = K$. In this simple case $K$ is limited to Mersenne numbers with the extension found in Ref.\cite{Palmer2021} allowing for an arbitrary integer $K$. Throughout the rest of this paper we refer to $K$ as the \emph{resolution} as this controls the minimum increment size of investments. The final cost function optimized on the quantum annealer is

\begin{align}
    \arg &\min_{\omega} -\omega^\mathsf{T}r + \gamma\left( \omega^\mathsf{T}\Sigma\omega\right) + ({\bf{1}}\cdot \omega - 1)^2
\end{align}

\noindent where the final term enforces the fully invested portfolio with no leverage constraint. More detailed descriptions on the implementation of quantum portfolio optimization can be found in Refs.\cite{Rosenberg2016, Mugel2020}.

\section{Cardinality Constraint}
\label{card}

\subsection{Description}

Cardinality constraints are used to limit the number of assets in a portfolio; the decision to use these constraints can be driven by reducing  management costs, transaction costs, or portfolio complexity, or by other investor preferences. One popular application is in financial index tracking where the objective is to replicate the behavior of a large set of assets with a limited subset \cite{Chang2000, Beasley2003}.

The introduction of the cardinality constraint makes the portfolio optimization problem non-convex due to the discrete decision of investing in an asset. The cardinality constrained Markowitz portfolio optimization can be written as

\begin{align}\label{eq:card_mark}
    \arg \min_{z,\omega} - &\left(Z \omega\right)^\mathsf{T} r + \gamma \left(Z\omega\right)^\mathsf{T}\Sigma\left(Z\omega\right),  \\ \nonumber 
    Z &= \text{diag}(z) \text{ ; } z \in \mathbb{Z}_2^N \\ \nonumber
    \text{subject to :} \\ \nonumber
    z^\mathsf{T}\omega &= 1 \\ \nonumber
    z^\mathsf{T}z &= C, \nonumber
\end{align}

\noindent where $z$ are binary indicator variables for assets that have been invested in ($z_i=1$) or not ($z_i=0$), the \emph{diag(.)} operator places these values along the diagonal of a zero matrix, and $C$ is the target portfolio cardinality value. Later on we shall discuss the application to cardinality-constrained index tracking portfolios which takes on the form of

\begin{align} \label{eq:tracking}
    \arg \min_{z,\omega} -\left( \left(Z\omega\right)^\mathsf{T} r - \hat{r}\right)^2 + \gamma \left(Z\omega\right)^\mathsf{T}\Sigma\omega. 
\end{align}

Due to the non-convex nature of the problem it cannot be solved using conventional convex optimization algorithms. Previous work involving cardinality-constraint optimization has primarily relied on the use of heuristic algorithms such as genetic algorithms \cite{Kim2019, Diaz2019}, or classical approximations \cite{Jiang2019, Graham2021} which do not scale well for large portfolios and are not practically reliable.

The difficulty of this problem when using classical optimization makes it an interesting case to study for quantum computing optimization, which is known to be able to solve complex non-convex optimization problems. The problem of cardinality-constrained quantum portfolio optimization and index tracking was recently addressed in Ref.\cite{Fernandez-Lorenzo2021} and Ref.\cite{Certo2022}. In the former it was proposed to use an iterative hybrid-classical quantum algorithm to find the discrete subset of assets. However, this approach does not solve the optimization problem directly and relies on multiple optimization procedures.
In the latter the cardinality constraint was formulated directly as QUBO for a quantum annealer but the formulation relied on an upper bound and was not exact which allowed for non-invested assets to be incorrectly identified as invested. 
The work presented here goes a step beyond the previous works by formulating the optimization problem in such a way that the cardinality-constrained optimization can be directly and exactly satisfied in a single optimization and is solved using quantum annealing. 

\subsection{Encoding}

The non-linear cardinality constraint needs to be formulated as a quadratic problem that can then be efficiently solved using a quantum annealer. When expanding Equations \ref{eq:card_mark} and \ref{eq:tracking} with respect to the problem variables $z_i$ and $\omega_i$ third and forth order terms arise. These high-order terms cannot be efficiently encoded on a quantum annealer. 

To enforce the cardinalty constraint of a value $C$ we add the additional term to minimise

\begin{align}
    \min_{z} \left({\bf{1}}\cdot z - C \right)^2
\end{align}

\noindent to the portfolio optimization objective function, where $z$ is the vector of asset indicator variables. We now need to formulate the design of the indicator variables $z_i$ for each asset.

\begin{figure}
    \centering
    \includegraphics[width=0.5\textwidth]{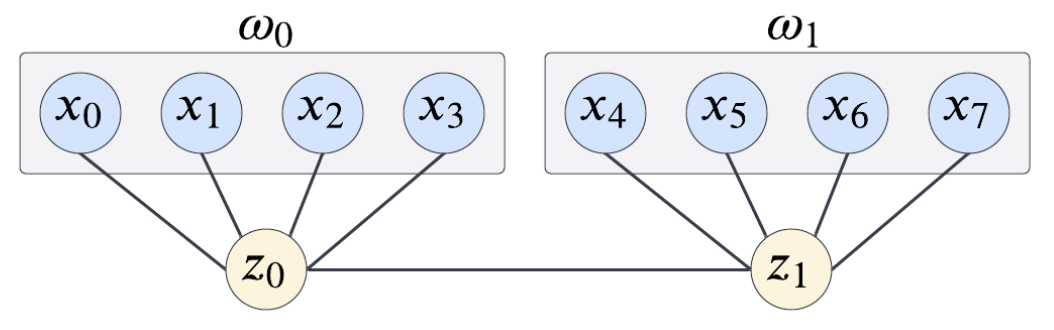}
    \caption{[Color online] Qubit interactions for the cardinality constraint implementation.}
    \label{fig:interactions}
\end{figure}

We can take advantage of the discretized portfolio optimization problem and apply indicator variables, $z_i$, with interactions between the discrete holding variables as shown in Figure \ref{fig:interactions}. Classically this approach would be avoided because of the NP-hardness; however, taking advantage of the power of the quantum optimizers now makes this a feasible approach.

It is worthwhile noting that for the cardinality constraint to be viable we require

\begin{align}
    \frac{K}{C} \geq 1,
\end{align}

\noindent otherwise it is not possible to have $C$ number of assets with minimum holding values to satisfy the total holding and cardinality constraint. From this we can also see that any active holding in a valid portfolio is bounded by
\begin{align}
     \frac{1}{K} \leq \omega_i \leq \frac{K - C +   1}{K},
\end{align}

\noindent which we can use to further reduce the range of integers required to encode the holdings with $K_{\max} = K - C + 1$. We can also see from this that we need to be careful when considering the value of $K$ to allow for upper bounds on the holdings. It is required that $\inf \sup \hat{\omega_i} \leq K_{\max}$, this implies that it is required that $ \lfloor \frac{K}{C} \rfloor \leq K_{\max}$.

\section{Index Tracking}
\label{ind}

Index tracking optimization aims to replicate the behavior of a target financial index of assets \cite{Beasley2003}, such as the Nasdaq-100 or S\&P 500 indexes studied here, using a tracking portfolio built from a small subset of the index assets. To track the index it is required that the returns and volatility of the tracking portfolio match as closely as possible. Simply measuring the deviation of the tracking portfolio returns with respect to the target index over a historical period provides a good tracking-error metric and gives robust holdings for out-of-sample tracking \cite{Beasley2003}

\begin{align}
    \epsilon_{TE} = \sum_t^T (\omega^\mathsf{T} r_t - \hat{r}_t)^2,
\end{align}

\noindent where $\omega$ is the tracking portfolio holdings, $r_t$ are the asset returns and $\hat{r}_t$ is the target index return. The cardinality constrained index tracking problem has been previously defined in Equation \ref{eq:tracking} and will be used as the optimization objective function. The full expression being minimized is 

\begin{align}\label{eq:final-tracking}
    \arg \min_{\omega, z} &\sum_t^T (\omega^\mathsf{T} r_t - \hat{r}_t)^2 + ({\bf{1}}\cdot\omega - 1)^2 \\ \nonumber
    &+ ({\bf{1}}\cdot z - C)^2 + \sum_{i,d} z_ix_{i,d},
\end{align}

\noindent where the final double summation terms are for the cardinality constraint indicator variable interaction terms. 

The quality of the tracking portfolios are measured with respect to the cumulative return series rather than the individual timestep returns. This measure ensures that the total return of the portfolio from the initial holding time is tracked, and that small errors in the timestep returns do not excessively compound, which can result in drift from the index returns over time. The \emph{cumulative return tracking error} (CTE) is defined as 

\begin{align}
    \epsilon_{CTE} = \sum_t^T \left[ \sum_{ 0\leq t^* \leq t} \log(1 + \omega^\mathsf{T} r_{t^*}) - \log(1 + \hat{r}_{t^*}) \right]^2 ,
\end{align}

where we have used the additive property of log returns.

\subsection{Data and Setup}

\begin{table}
   % \centering
\begin{tabular}{rrrrrr}
\hline
\multicolumn{1}{l}{N} & \multicolumn{1}{l}{K} &
\multicolumn{1}{l}{Max \% Holding} & \multicolumn{1}{l}{$K_{\max}$} & \multicolumn{1}{l}{D} & \multicolumn{1}{l}{Total Qubits} \\ \hline
100                        & 31                             & 20\%                               & 6                               & 3                             & 400                            \\
100                        & 63                             & 20\%                               & 12                              & 4                             & 500                            \\
100                        & 127                            & 20\%                               & 25                              & 5                             & 600                            \\
100                        & 255                            & 20\%                               & 51                              & 6                             & 700                            \\
500                        & 127                            & 20\%                               & 25                              & 5                             & 3000 \\ \hline \hline                         
\end{tabular}
    \caption{Problem size given by the number of qubits. The max holding is the rounded integer \% of the resolution, and the bit depth is calculated as the maximum number of bits required in a binary encoding scheme to represent the max holding value. The total qubits is given by $N(D + 1)$, where the $+1$ corresponds to the additional cardinality constraint variables.}
    \label{tab:qubits}
\end{table}

In this study the financial data used are daily returns for the Nasdaq-100 (ticker NDK) and S\&P 500 (ticker GSPC). The data is taken over the period of JUN-01-2021 $\rightarrow$ MAY-28-2022.

To aid the optimization the number of qubits can be reduced by defining upper bounds on the asset holdings and using the method in Ref.\cite{Palmer2021} for the asset encoding. In this instance we have allowed a single asset to have a maximum holding of 20\% in the portfolio; for comparison the Nasdaq-100 index has a maximum holding of 13\%. Table \ref{tab:qubits} summarizes the number of qubits required using the holding bounds. Where $K$ is the total number of units of investment available to allocate, the maximum holding $K_{\max}$ is then the rounded integer percentage of the resolution. 

For large numbers of assets this technique can lead to considerable savings in resources where the unbounded problem size, $K_{\max} = K$, would be too large to use with the current quantum solvers which at the time of writing are capped at around 4000 \cite{dwave}.

To solve the QUBO optimization problem we are using the D-Wave LEAP Hybrid solver.

\subsection{Results}

\begin{table}[]
\centering
\footnotesize
\begin{tabular}{ccccccc}
\multicolumn{6}{c}{Best Portfolios Nasdaq-100}                                                                                                                                                                 \\ \hline 
\multicolumn{1}{c}{C}  & \multicolumn{1}{c}{K}                                      & $\epsilon_{CTE}$              & MRE                       & MdRE                      & Vol Error                  \\ \hline
25                    & 31                                                      & \multicolumn{1}{c}{0.00024} & \multicolumn{1}{c}{0.280} & \multicolumn{1}{c}{0.117} & \multicolumn{1}{c}{7.00\%} \\
25                    & 63                                                       & \multicolumn{1}{c}{0.00012} & \multicolumn{1}{c}{0.205} & \multicolumn{1}{c}{0.071} & \multicolumn{1}{c}{2.00\%} \\
25                    & 127                                                      & \multicolumn{1}{c}{0.00031} & \multicolumn{1}{c}{0.263} & \multicolumn{1}{c}{0.175} & \multicolumn{1}{c}{1.00\%} \\ \hline
50                    & 63                                                      & \multicolumn{1}{c}{0.00016} & \multicolumn{1}{c}{0.205} & \multicolumn{1}{c}{0.106} & \multicolumn{1}{c}{4.00\%} \\
50                    & 127                                                       & \multicolumn{1}{c}{0.00017} & \multicolumn{1}{c}{0.141} & \multicolumn{1}{c}{0.115} & \multicolumn{1}{c}{7.00\%} \\ 
50                    & 255                                                       & \multicolumn{1}{c}{0.00027} & \multicolumn{1}{c}{0.233} & \multicolumn{1}{c}{0.111} & \multicolumn{1}{c}{4.00\%} \\ \hline
75                    & 127                                                       & \multicolumn{1}{c}{0.00012} & \multicolumn{1}{c}{0.199} & \multicolumn{1}{c}{0.069} & \multicolumn{1}{c}{1.00\%} \\
75                    & 255                                                       & \multicolumn{1}{c}{0.00033} & \multicolumn{1}{c}{0.229} & \multicolumn{1}{c}{0.182} & \multicolumn{1}{c}{1.00\%} \\ \hline \hline

\multicolumn{6}{c}{Best Portfolios S\&P 500} \\
\hline
\multicolumn{1}{l}{} & \multicolumn{1}{l}{} & \multicolumn{1}{l}{$\epsilon_{CTE}$} & \multicolumn{1}{l}{MRE} & \multicolumn{1}{l}{MdRE} & \multicolumn{1}{l}{Vol Error} \\ \hline
50                    & 63                             & 0.00021                                & 0.815                        & 0.182                         & 0.08\%                             \\
100                   & 255                            & 0.00021                                & 0.945                        & 0.223                         & 0.05\% \\ \hline \hline                           
\end{tabular}
\caption{Index tracking results for the Nasdaq-100 index. MRE is the mean relative error taken over all samples results, and MdRE is the median relative error.}
\label{tab:tracking-results}
\end{table}

\begin{table}[]
\centering
\begin{tabular}{c|ccc} 
\multicolumn{1}{l}{} & \multicolumn{3}{c}{$C$} \\ \hline
$K$                    & 25     & 50    & 75   \\ \hline
31                   & 70\%   & n/a   & n/a  \\
63                   & 100\%  & 80\%  & n/a  \\
127                  & 100\%  & 90\%  & 40\% \\
255                  & n/a    & 100\% & 80\% \\ \hline \hline
\end{tabular}
\caption{Success rates of the 20 D-Wave samples for finding feasible solutions that satisfy all of the problem constraints. $C$ is the problem cardinality, $K$ is the resolution. A value of n/a indicates the experiment was not conducted.}
\label{tab:success}
\end{table}

\begin{figure*}[t]
\resizebox{\textwidth}{!}{
    \centering
    \begin{subfigure}[b]{0.3\textwidth}
    \centering
    \includegraphics[width=\textwidth]{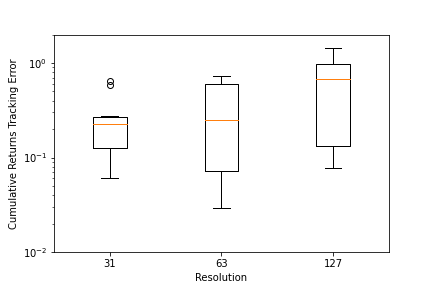}
    \caption{$C=25$}
    \end{subfigure}
 
    \begin{subfigure}[b]{0.3\textwidth}
    \centering
    \includegraphics[width=\textwidth]{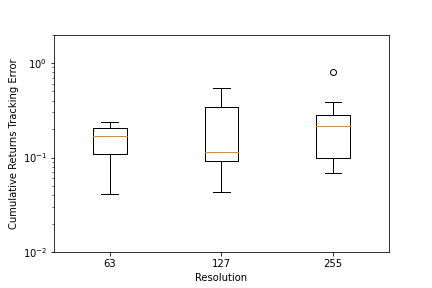}
    \caption{$C=50$}
    \end{subfigure}

    \begin{subfigure}[b]{0.3\textwidth}
    \centering
    \includegraphics[width=\textwidth]{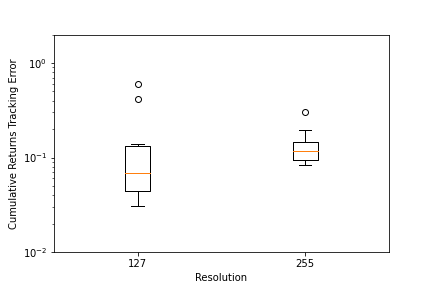}
    \caption{$C=75$}
    \end{subfigure}
    }
    \caption{[Color online] Boxplots showing the distribution of the Nasdaq-100 cumulative returns tracking errors for sets of cardinality constraints $C$ and resolutions $K$; 20 sample portfolios were obtained from the D-Wave Leap Hybrid optimizer for each set of parameters.}
    \label{fig:boxplot}
\end{figure*}

\begin{figure}
\centering
    \begin{subfigure}[b]{0.4\textwidth}
    \includegraphics[width=\textwidth]{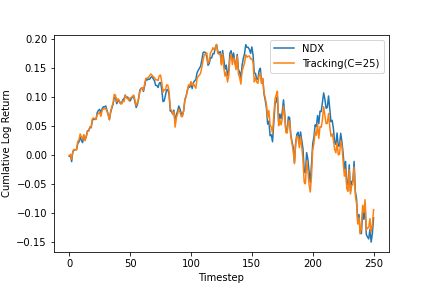}
    \end{subfigure}
    \begin{subfigure}[b]{0.4\textwidth}
    \includegraphics[width=\textwidth]{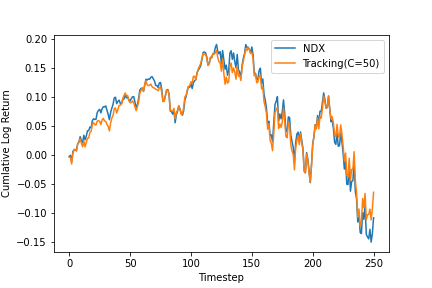}
    \end{subfigure}
    \begin{subfigure}[b]{0.4\textwidth}
    \includegraphics[width=\textwidth]{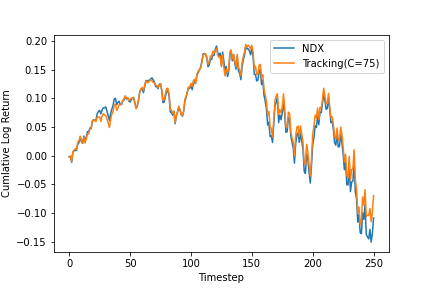}
    \end{subfigure}
    \caption{[Color online] Best tracking portfolios for Nasdaq-100 for different values of the cardinality constraints C=25, 50, and 75 respectively.}
    \label{fig:ndx-tracking}
\end{figure} 

\begin{figure}
    \centering
    \includegraphics[width=0.4\textwidth]{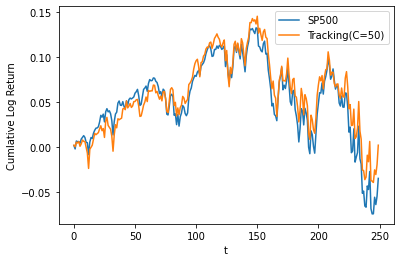}
    \caption{[Color online] S\&P 500 tracking portfolio with $C=50$.}
    \label{fig:sp500-tracking}
\end{figure}

\begin{figure*}
\resizebox{\textwidth}{!}{
    \centering
    \includegraphics[width=\textwidth]{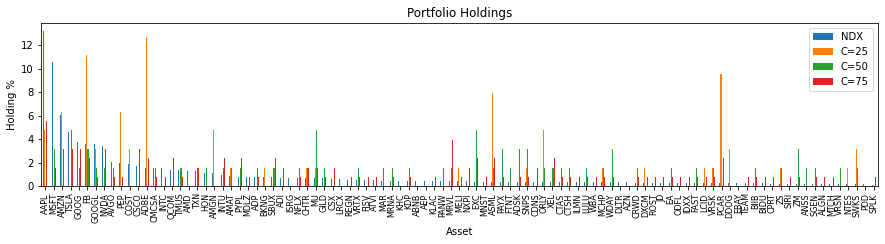}
}
    \caption{[Color online] Nasdaq-100 index (NDX) best tracking portfolio compositions for cardinalities $C=25, 50, 75$.}
\label{fig:ndx-composition}
\end{figure*}

For the index tracking problem various cardinalities, $C$, and holding resolutions, $K$, are investigated. Table \ref{tab:tracking-results} summarizes the optimization results for the Nasdaq-100 and S\&P500 tracking, showing the metrics for the best solution in each case. The mean relative error (MRE) and median relative error (MdRE) are taken over the cumulative tracking errors. Both the mean and median are used because the mean can be skewed by a large relative error occurring in only a small section of the tracking; in these examples this can be observed to sometimes occur near the end of the timeseries.

We used 20 samples taken from the D-Wave Leap Hybrid optimizer. Firstly we can see from Table \ref{tab:success} that the success rate of finding feasible portfolios is very high, being close to 100\% in most cases. This indicates that our formulation of the cardinality constraint is extremely effective and reliable for finding feasible solutions. 

For each of the sets of $C$ and $K$ explored Figure \ref{fig:boxplot} shows the distribution of the cumulative tracking errors for these samples. As the cardinality increases the distribution of tracking errors decreases, which can be explained by the availability of more assets naturally being able to represent a higher proportion of the target index behavior. There is a slight trend that as the resolution increases the upper quantiles of the sample tracking errors increase which is due to the size of the search space increasing. We can see there is a balance between having a low enough resolution such that the search space does not become too large, and high enough such that it is expressible enough for good solutions.

The two best tracking error portfolios, $\epsilon_{CTE}=0.00012$ and with very low volatility errors, were for $C=25$ with resolution$=63$, and $C=75$ with resolution$=127$. It could have been expected that $C=75$ would easily give the best tracking portfolio, these results importantly demonstrate the relevance of cardinality constraints such that small portfolios can deliver the same performance. 

In Figure \ref{fig:ndx-composition} we compare the holdings of the best tracking portfolios for each cardinality. We can observe a large imbalance in the portfolio holdings for $C=25$, with some interesting heavy weightings on low weighted index assets. This behavior occurs because smaller portfolios rely on a specific asset to represent the behavior of a larger set of correlated assets. It is also this imbalance that results in the higher deviation of sample results in Figure \ref{fig:boxplot}. For the the higher cardinalities the holdings become more uniform as this becomes a closer approximation to the actual index weights, and as has been observed this can be better approximated by lower resolutions.

We finally apply this methodology to track the 500 asset S\&P 500 index; this pushes the limits of the optimization with respect to problem size, see Table \ref{tab:qubits}. Even though the problem is now extremely large we are still able to obtain very good tracking portfolios with tracking errors similar to the 100 asset case. Although the mean relative errors are rather large and close to 100\% error in the returns we can see from Figure \ref{fig:sp500-tracking} that the majority of the tracking fits very well and is better reflected by the median relative error, which is about 20\%; considering the respectively smaller sizes of the tracking portfolios this is very encouraging. The small errors in the volatility, both with only an excess of 0.01\% of risk, also support the quality of both tracking portfolios.

These results show that we are able to successfully use quantum optimization to replicate large financial indexes with only a small subset of assets; this has important implications in the development of financial products such as exchange traded funds (ETFs).

\section{Enhanced Index Tracking}
\label{enh}

\begin{figure*}
\resizebox{\textwidth}{!}{
    \centering
    \includegraphics[width=\textwidth]{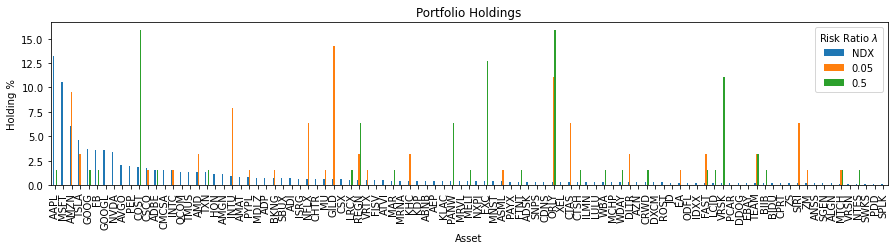}
}
    \caption{[Color online] Nasdaq-100 index (NDX) enhanced tracking portfolio compositions for 25 assets with risk ratios $\lambda=0.05$ which gives the best enhancement score, and $\lambda=0.5$ which gives the best Sharpe ratio improvement.}
    \label{fig:enhanced-ndx-composition}
\end{figure*}

As a progression from index tracking the objective of enhanced index tracking is to create a portfolio that tracks the index behavior closely, but not exactly, in order to create more beneficial investment preferences over the target index. Typical index-beating ETFs rely on leverage, which is not used here and we instead rely on an optimized allocation of the index assets.

The standard financial indexes tend to be constructed with respect to representation rather than optimized for investment metric performance; for example the assets in an index may be weighted by market capitalization or other non-technical factors. An enhancement which may be desired is, for example, improving the risk-return profile by finding a good tracking portfolio but with lower covariance compared to the target index; this is the approach taken here.

The expression to minimize is Equation \ref{eq:final-tracking} with the additional covariance term $\omega^\mathsf{T}\Sigma\omega$. The first term in Equation \ref{eq:final-tracking} now becomes
    
    \begin{align}
    \sum_t^T (1 - \lambda)\left(\omega^\mathsf{T} r_t - \hat{r}_t\right)^2 + \lambda\left(\omega^\mathsf{T} r_t   + \gamma\omega^\mathsf{T}\Sigma_t \omega\right).
    \end{align}
    
The risk-ratio parameter $\lambda$ is used to control the ratio of the Markowitz portfolio optimization cost and tracking cost terms; when $\lambda=0$ then there is no enhancement. This balances between finding a tracking portfolio with finding the best mean-variance portfolio. 

\subsection{Data and Setup}

In this Section we focus on tracking portfolio optimization constrained to 25 assets and using a resolution of $K=63$. The lower number of assets results in more interesting portfolio compositions, and a resolution of 63 gives the best performing index tracking portfolio as demonstrated in previous results. 
The risk-ratio parameter is varied over logarithmic scale to examine the impact of both extremes of values of $\lambda$. 
The covariance matrices, $\Sigma_t$, for each timestep were calculated using a 90-day rolling window.

\subsection{Results}

\begin{table}[]
\centering
\footnotesize
\begin{tabular}{cccccc}
\multicolumn{6}{c}{Best Portfolios} \\ \hline
\multicolumn{1}{l}{Ratio} & \multicolumn{1}{l}{$\epsilon_{CTE}$} & \multicolumn{1}{l}{Vol Error} & \multicolumn{1}{l}{MDRSE} & \multicolumn{1}{l}{Correlation} & \multicolumn{1}{l}{Sharpe/$\epsilon_{CTE}$} \\ \hline
0.00                      & 0.00018                            & -3.00\%                                & 0.04                                      & 0.96                            & 2.67                                      \\
0.05                      & 0.00030                            & -20.86\%                               & 0.36                                      & 0.94                            & 20.65                                     \\
0.10                      & 0.00188                            & -27.68\%                               & 0.70                                      & 0.91                            & 16.17                                     \\
0.20                      & 0.00537                            & -26.13\%                               & 0.85                                      & 0.86                            & 11.58                                     \\
0.50                      & 0.01688                            & -21.32\%                               & 1.26                                      & 0.89                            & 9.73                                      \\
\hline \hline                                  
\end{tabular}
\caption{Metrics for the best Nasdaq-100 enhanced tracking portfolios with different covariance minimization ratios. The best Sharpe/$\epsilon_{CTE}$ is taken from 10 samples using D-Wave Leap Hybrid. MDRSE = the median relative Sharpe ratio error with respect to the target index timestep Sharpe ratios. The correlation is measured between the timestep returns, not the cumulative as in the CTE.}
\label{tab:enhanced-results}
\end{table}

\begin{figure}
    \centering
    \begin{subfigure}[b]{0.49\textwidth}
    \centering
    \includegraphics[width=\textwidth]{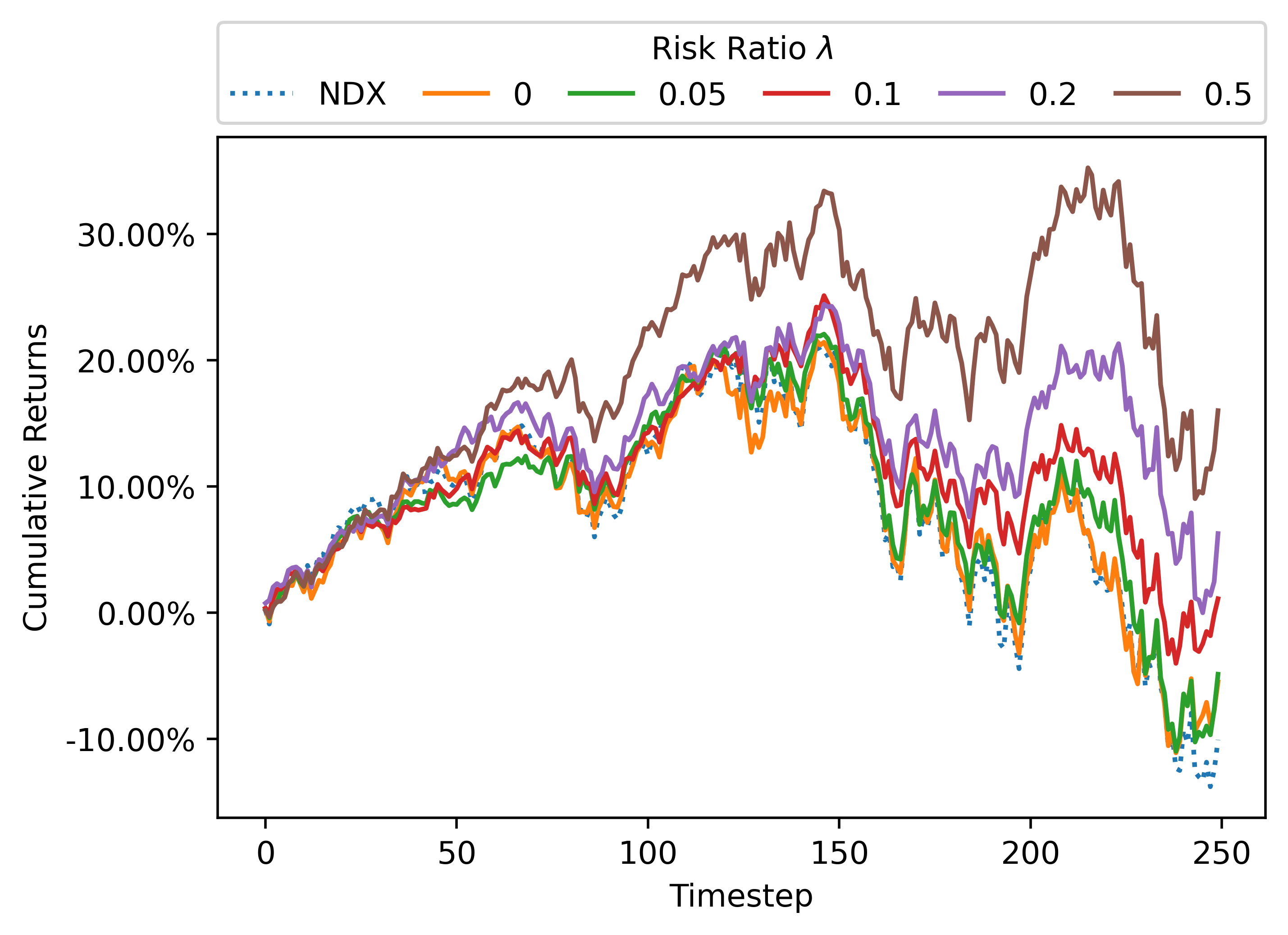}
    \end{subfigure}
    \begin{subfigure}[b]{0.49\textwidth}
    \centering
    \includegraphics[width=\textwidth]{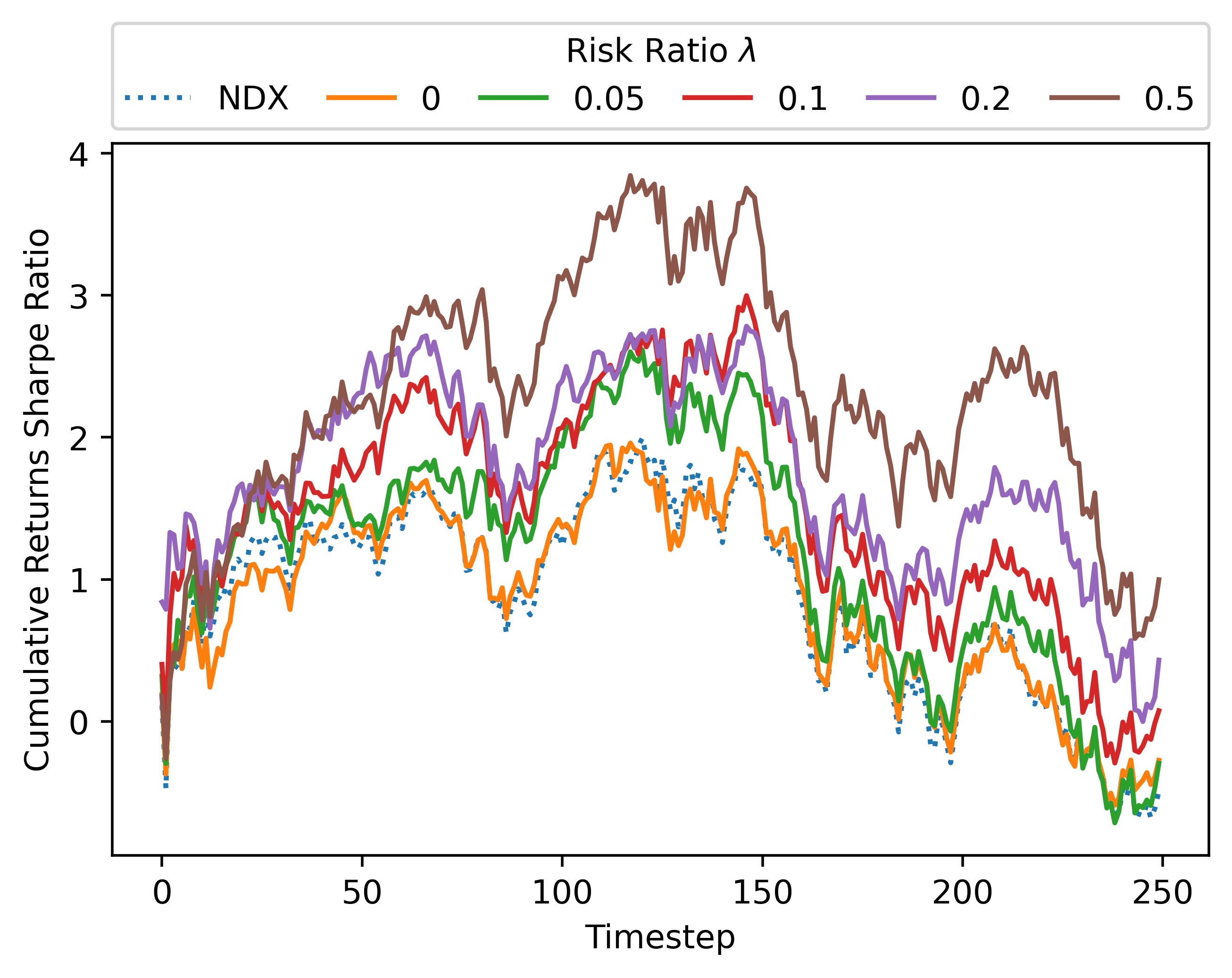}
    \end{subfigure}
    \caption{[Color online] Enhanced Nasdaq-100 (NDX) tracking portfolios with $C=25$ and varying covariance minimization ratios 0.2, 0.5 and 0.95: cumulative returns (upper panel) and cumulative returns Sharpe ration (lower panel).}
    \label{fig:enhanced-tracking}
\end{figure}

\begin{figure}
    \centering
    \includegraphics[width=0.49\textwidth]{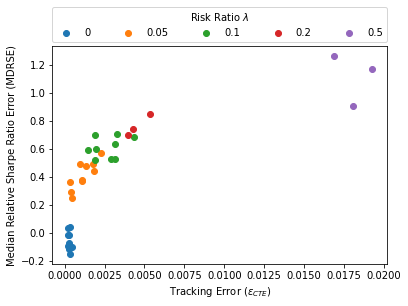}
    \caption{[Color online] Tracking error vs relative Sharpe ratio error for enhanced tracking portfolio samples. The ratio of the covariance minimization objective term was varied.}
    \label{fig:sharpe_vs_tracking}
\end{figure}

The risk-ratio $\lambda$ parameter is varied to explore how the contribution of the mean-variance minimization affects the tracking and provides enhancement to the target index. To measure the enhanced portfolio we use the enhancement score as the ratio between the improvement of the Sharpe ratio, where the Sharpe ratio is given by $\frac{w_t^\mathsf{T}r_t}{\sqrt{w_t^\mathsf{T}\Sigma_tw_t}}$, and cumulative return tracking error. This is given in the last column of Table \ref{tab:enhanced-results}.

The improvement of the Sharpe ratio is measured as the median relative error over the timesteps of the tracking Sharpe ratio compared to the target index Sharpe ratio. In this application, a large positive error in the Sharpe ratio is beneficial as it indicates a more favorable Sharpe ratio of the tracking portfolio; similarly, a negative volatility error indicates a more beneficial risk profile.

The full set of metrics for the best tracking portfolios, determined by the highest enhancement score, for each of the risk-ratios considered are given in Table \ref{tab:enhanced-results}.
Figure \ref{fig:enhanced-tracking} shows the cumulative tracking returns and Sharpe ratio of the best enhanced portfolios compared to the target index. As expected, when $\lambda=0$ the portfolio tracks the target index very closely in both returns, a 96\% correlation, and Sharpe ratio, with only a $-3.00\%$ improvement in volatility; which gives the lowest enhancement score. However, after introducing a small amount of mean-variance optimization when $\lambda=0.05$ we already see a significant improvement in the risk profile with a 20\% improvement in volatility, resulting in a median Sharpe ratio improvement of 36\%. Incredibly, this is all whilst retaining a very low cumulative tracking error with 94\% timestep returns correlation to the target index; in Figure \ref{fig:enhanced-tracking} it can be seen to be very close to $\lambda=0$ in terms of tracking performance. The combined result of these observations gives $\lambda=0.05$ the highest enhancement score and demonstrates the extreme practical value of performing enhanced index tracking.

As the risk-ratio, $\lambda$, increases past $0.05$ the enhancement score begins to decrease which is due to the decrease in cumulative returns tracking. To improve the Sharpe ratio excess returns are generated which cause for a deviation from the target index. In Table \ref{tab:enhanced-results} there is a clear trend that the tracking error increases and returns correlation decreases as we increase $\lambda$. The volatility improvements stay within the 20-30\% range which indicates that the improvements in Sharpe ratio are from excess returns. When $\lambda=0.5$ the mean-variance optimization dominates and generates over 100\% (2x) improvement in the Sharpe ratio; even though the cumulative returns do not track well, there is still a 90\% correlation with the timestep returns. This demonstrates that significant improvements in the risk profile can be made if exact tracking is not the primary objective.

% Figure \ref{fig:enhanced-tracking} shows the cumulative tracking returns and Sharpe ratio of the three best enhanced portfolios compared to the target index. We can see, as expected, that the enhanced portfolios track the target index returns well but with significant improvements in the Sharpe ratio as $\lambda$ increases. For $\lambda=0.20$ we can achieve a well balanced enhanced portfolio with a low tracking error, high returns correlation of 91\% with the target index whilst still improving the Sharpe ratio. With the largest $\lambda=0.95$ there is up to a 2x improvement in the Sharpe ratio but compensating with a small reduction in the returns correlation; this is the most beneficial enhanced portfolio.

To gain more insight into the optimization behavior Figure \ref{fig:sharpe_vs_tracking} shows the tracking error vs the median relative Sharpe ratio error. We can see there is a clear trend that as the tracking error increases the improvement in the Sharpe ratio increases. It is interesting to see that this trend is a curve similar to the Markowitz efficient frontier and shows the relationship between risk and return. As was previously noted as $\lambda$ increases the Sharpe ratio improvement becomes more driven by excess returns whilst the volatility stays within a given range; from this figure we can see that the optimal enhancement scores are the regions of largest gradient which are primarily driven by improving the volatility. There is then a progressive trade-off between the objective of how well it is desired to track the index and the excess returns which is shown by the trend of the enhancement scores. 

The two sets of portfolio holdings for $\lambda=0.05,0.5$ are given in Figure \ref{fig:enhanced-ndx-composition}, it is worth noting that even though the behaviours as seen are very similar to the target index the compositions are vastly different. This suggests that a set of low covariance assets can be used to capture the target index returns behavior. One possible explanation is that the index is composed of assets representing different sectors and assets in different sectors are more likely to have lower covariance with each other. To track the index behavior well each of the sectors are required to be reasonably well represented by the limited selection of assets. Therefore a small subset of assets in each of the sectors would likely give a relative low covariance portfolio and diversification which is encouraged by larger risk-ratio optimizations.
    
% For high Sharpe ratios there is not much of a clear trend with respect to the tracking errors except that they cluster around the area of moderately good tracking errors between 0.00200 - 0.00400. This suggests that the exact behavior of the target index does not exhibit optimal risk, however large deviations from the target index are not required to significantly enhance performance.
    
% The larger risk-ratios, $\lambda$, give the best Sharpe ratios as expected, but unexpectedly the majority have good tracking errors. This further suggests that a set of low covariance assets can be used to capture the target index returns behavior. One possible explanation is that the index is composed of assets representing different sectors and assets in different sectors are more likely to have lower covariance with each other. To track the index behavior well each of the sectors are required to be reasonably well represented by the limited selection of assets. Therefore a small subset of assets in each of the sectors would likely give a relative low covariance portfolio and diversification which is encouraged by larger risk-ratio optimizations.
    
We have shown that when using quantum computing for non-linearly constrained optimization, with cardinality constraints it is possible to create smaller and better performing portfolios whilst retaining a high degree of similarity to popular financial indexes.

\section{Conclusion and Outlook}

In this work we have shown how non-linear portfolio cardinality constraints can be implemented when solving the portfolio optimization problem with a quantum annealer, in the context of index tracking. By taking advantage of the discrete portfolio optimization it was possible to formulate the constrained problem with at most quadratic terms, an approach which would otherwise be unfeasible for classical methods. We then use D-Wave's Hybrid quantum annealing to solve the optimization problem.

This methodology was applied to index tracking and enhanced index tracking for the Nasdaq-100 and S\&P 500 indexes. We were able to construct high quality tracking portfolios to capture the target index returns using significantly smaller subsets of assets, 4x and 10x smaller for the Nasdaq-100 and S\&P 500 respectively. For enhanced index tracking we were able to construct tracking portfolios that significantly outperformed the risk profile of the target index by up to 2x whilst retaining a high degree of correlation with the returns of the target index.

Examples provided in this work are initial practical applications of the index tracking problems to test our formulation of the cardinality constraint. Future work would be to further enhance the complexity of the index tracking model, and include additional terms such as dividend reinvestment, transaction costs for re-balancing events, and other types of assets used for building advanced ETF products. Most importantly, we have shown how we can apply current quantum computers to a real-world practical problem in finance, and hope that this work encourages the adoption of quantum computing as a tool that can provide real value today.

\bibliographystyle{apsrev4-2}
\bibliography{ally}

\end{document}